\documentclass[journal,letterpaper]{IEEEtran_nssmic}
\usepackage{xspace,graphicx} 

\usepackage{amsmath}  
\input colordvi.tex

\begin{document}

\def\BF{$B$ Factory\xspace}
\def\abf {asymmetric \BF}
\def\B    {\ensuremath{B}\xspace}
\def\piz  {\ensuremath{\pi^0}\xspace}
\def\epem {\ensuremath{e^+e^-}\xspace}
\def\en   {\ensuremath{e^-}\xspace}   
\def\ep   {\ensuremath{e^+}\xspace}
\def\Bz   {\ensuremath{B^0}\xspace}
\newcommand{\nimBaseC}{Nucl.\ Instr.\ and Methods\xspace}
\newcommand{\nima}[1]{\nimBaseC~A~{\bf #1}}
\newcommand{\gevcc}{\ensuremath{{\mathrm{\,\hbox{Ge}\kern -0.1em \hbox{V}\!/}c^2}}\xspace}

\def\titlebabar{\mbox{\slshape B\kern-0.1em{\huge A}\kern-0.1em
    B\kern-0.1em{{\huge A}\kern-0.1em {\huge R}}}\xspace} 
\def\babar{\mbox{\slshape B\kern-0.1em{\footnotesize A}\kern-0.1em
    B\kern-0.1em{\footnotesize A\kern-0.1em R}}\xspace}
\def\referencebabar{\mbox{\slshape B\kern-0.1em{\scriptsize A}\kern-0.1em
    B\kern-0.1em{\scriptsize A\kern-0.2em R}}\xspace}
\def\etal{\hbox{\em et al.\null}}
\newcommand{\GeV}{\ensuremath{\,}Ge\ensuremath{\kern-0.1em}V\xspace}
\newcommand{\MeV}{\ensuremath{\,}Me\ensuremath{\kern-0.1em}V\xspace}
\newcommand{\MeVnospace}{Me\ensuremath{\kern-0.1em}V\xspace}
\newcommand{\KeV}{\ensuremath{\,}ke\ensuremath{\kern-0.1em}V\xspace}
\newcommand{\boldgev}{\ensuremath{\,}Ge\ensuremath{\kern-0.1em}V\xspace}
\newcommand{\boldmev}{\ensuremath{\,}Me\ensuremath{\kern-0.1em}V\xspace}
\newcommand{\boldkev}{\ensuremath{\,}ke\ensuremath{\kern-0.1em}V\xspace}
\newcommand{\boldBF}{{\bf B} Factory\xspace}
\newcommand{\boldabf}{asymmetric {{\bf B} Factory}\xspace}
\newcommand{\picwid}{3.5in}
\newcommand{\hz}{\ensuremath{\,}Hz\xspace}
\newcommand{\khz}{\ensuremath{\,}kHz\xspace}
\newcommand{\mhz}{\ensuremath{\,}MHz\xspace}
\newcommand{\mmeter}{{\ensuremath{\,}}mm\xspace}
\newcommand{\meter}{{\ensuremath{\,}}m\xspace}
\newcommand{\nsec}{{\ensuremath{\,}}ns\xspace}
\newcommand{\msec}{{\ensuremath{\,}}ms\xspace}
\newcommand{\usec}{{\ensuremath{\,\mu}s}\xspace}
\def\y#1{{#1}}

\title{ Absolute Energy Calibration with the \\ Neutron-Activated 
    Liquid-Source System \\ at \titlebabar's CsI(Tl) Calorimeter}

\author{Johannes M. Bauer,~\IEEEmembership{Member,~IEEE},~%
{\em for the EMC group of the \babar Collaboration}%
\thanks{Manuscript received November 14, 2003.  
        The author was supported by  U.S. Dept. of 
        Energy grant DE-FG05-91ER40622.}%
\thanks{J. M. Bauer is with the Department of Physics and Astronomy, 
        University of Mississippi, University, MS 38677, USA
        (e-mail: bauerj@slac.stanford.edu).}}

\newif\ifslacpub  \slacpubtrue

\ifslacpub
\onecolumn

\renewcommand{\thefootnote}{\fnsymbol{footnote}}
\def\babar{\mbox{\slshape B\kern-0.1em{\footnotesize A}\kern-0.1em
    B\kern-0.1em{\footnotesize A\kern-0.1em R}}\xspace}
\newcommand{\mev}{\ensuremath{\,}Me\ensuremath{\kern-0.1em}V\xspace}
\newcommand{\gev}{\ensuremath{\,}Ge\ensuremath{\kern-0.1em}V\xspace}

\begin{flushright}
{\small
SLAC-PUB-10289\\
UMS-HEP-2003-027\\
November 2003\\}
\end{flushright}

\vspace{.8cm}

\begin{center}
{\large\bf Absolute Energy Calibration with the \\ Neutron-Activated
Liquid-Source System \\ at \babar's CsI(Tl)
Calorimeter\footnote{Work supported by Department of Energy contract
DE--AC03--76SF00515 and Department of Energy grant DE-FG05-91ER40622.}\\}

\vspace{1cm}

J. M. Bauer\\
Department of Physics and Astronomy \\ 
University of Mississippi, \\
University, MS 38677, USA \\

\medskip
{\em for the EMC group of the \babar Collaboration}

\medskip
Stanford Linear Accelerator Center\\ 
Stanford University \\
Stanford, CA 94309, USA \\

\end{center}

\vfill

\begin{center}
{\large\bf
Abstract }
\end{center}

\begin{quote}
\leftskip25mm\rightskip25mm
The electro-magnetic calorimeter at the \babar detector, part of the
asymmetric B~Factory at SLAC, measures photons in the energy range
from 20\mev to 8\gev with good resolution.  The~calorimeter~is
calibrated at the low energy end with 6.13\mev photons obtained from a
liquid source system.  During the calibration, a fluorine-rich liquid is
activated via a neutron generator and pumped past the front of the
calorimeter's crystals.  Decays that occur in front of the crystals emit
photons of well-defined energy, which are detected in the crystals with
the regular data acquisition system.  The~liquid source system adds only
very little material in front of the calorimeter, needs nearly no
maintenance, and allows operation at the switch of a key with minimal
safety hazards.  The~report describes the system, presents calibration
results obtained from its operation since 1999, shows the crystals' loss
of light yield due to radiation damage, and shares experiences gained
over the years.
\end{quote}

\vfill

\begin{center}
{\it Presented at the 2003 IEEE Nuclear Science Symposium and Medical Imaging Conference \\
Portland, OR, USA \\
October 19, 2003 -- October 25, 2003 \\ [1.5ex]
Submitted to IEEE Transactions on Nuclear Science} \\
\end{center}

\twocolumn
\fi

\maketitle

\begin{abstract}
The electro-magnetic calorimeter at the \babar detector, part of the
\boldabf at SLAC, measures photons in the energy range from 20\boldmev
to 8\boldgev with good resolution.  The~calorimeter~is calibrated at the
low energy end with 6.13\boldmev photons obtained from a liquid source
system.  During the calibration, a fluorine-rich liquid is activated
via a neutron generator and pumped past the front of the calorimeter's
crystals.  Decays that occur in front of the crystals emit photons of
well-defined energy, which are detected in the crystals with the
regular data acquisition system.  The~liquid source system adds only
very little material in front of the calorimeter, needs nearly no
maintenance, and allows operation at the switch of a key with minimal
safety hazards.  The~report describes the system, presents calibration
results obtained from its operation since 1999, shows the crystals' loss
of light yield due to radiation damage, and shares experiences gained
over the years.
\end{abstract}


\section{Introduction}

\begin{figure}
\centering
\includegraphics[width=2.4in]{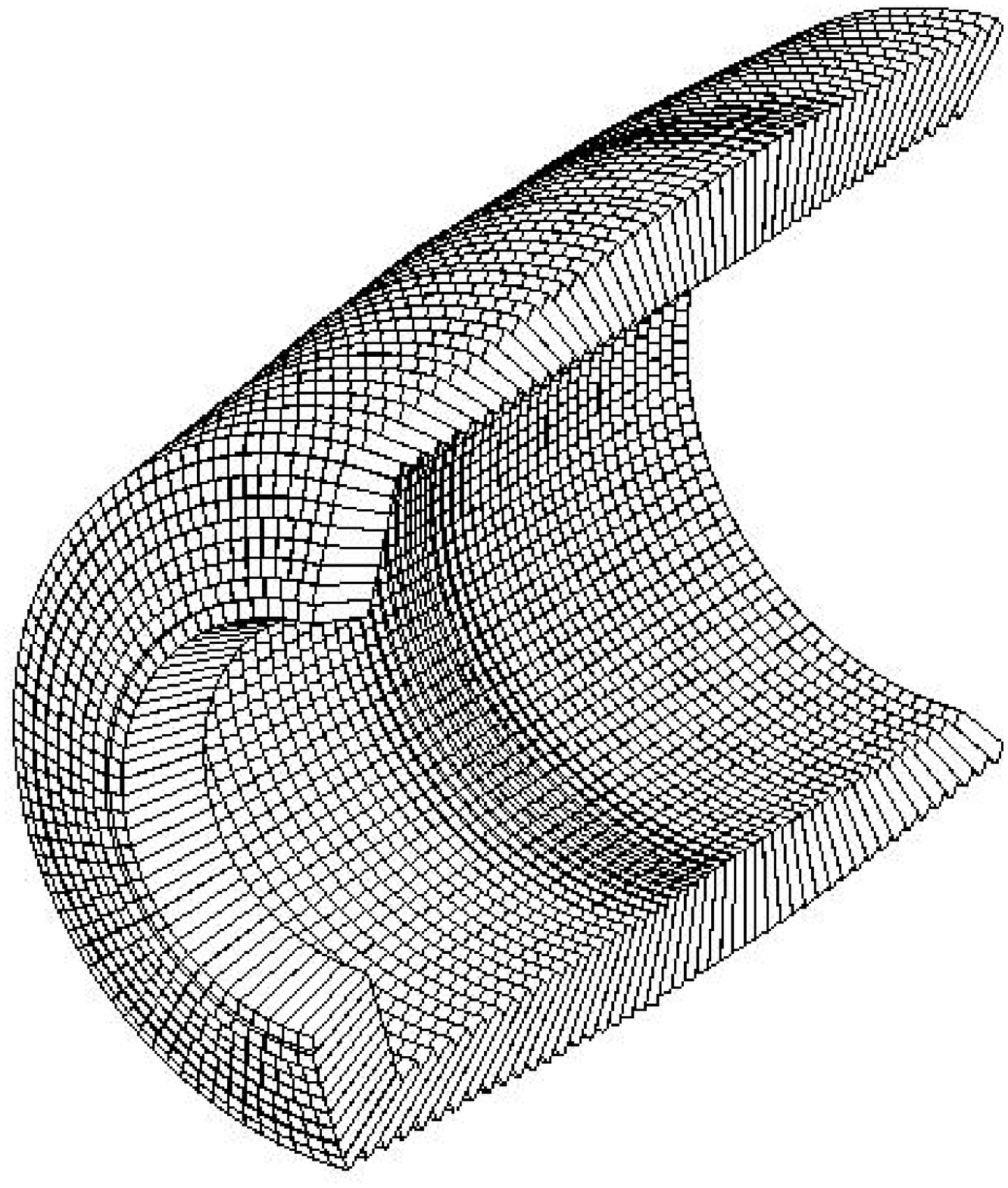}
\caption{Cut-out drawing of the cylindrical electromagnetic
calorimeter's arrangement of crystals.  The support structure (not
shown) is at the back of the crystals to minimize the amount of material
in front of the crystals.}\label{fig:gv_emccut}%
\vspace{3mm}
\includegraphics[width=3.5in]{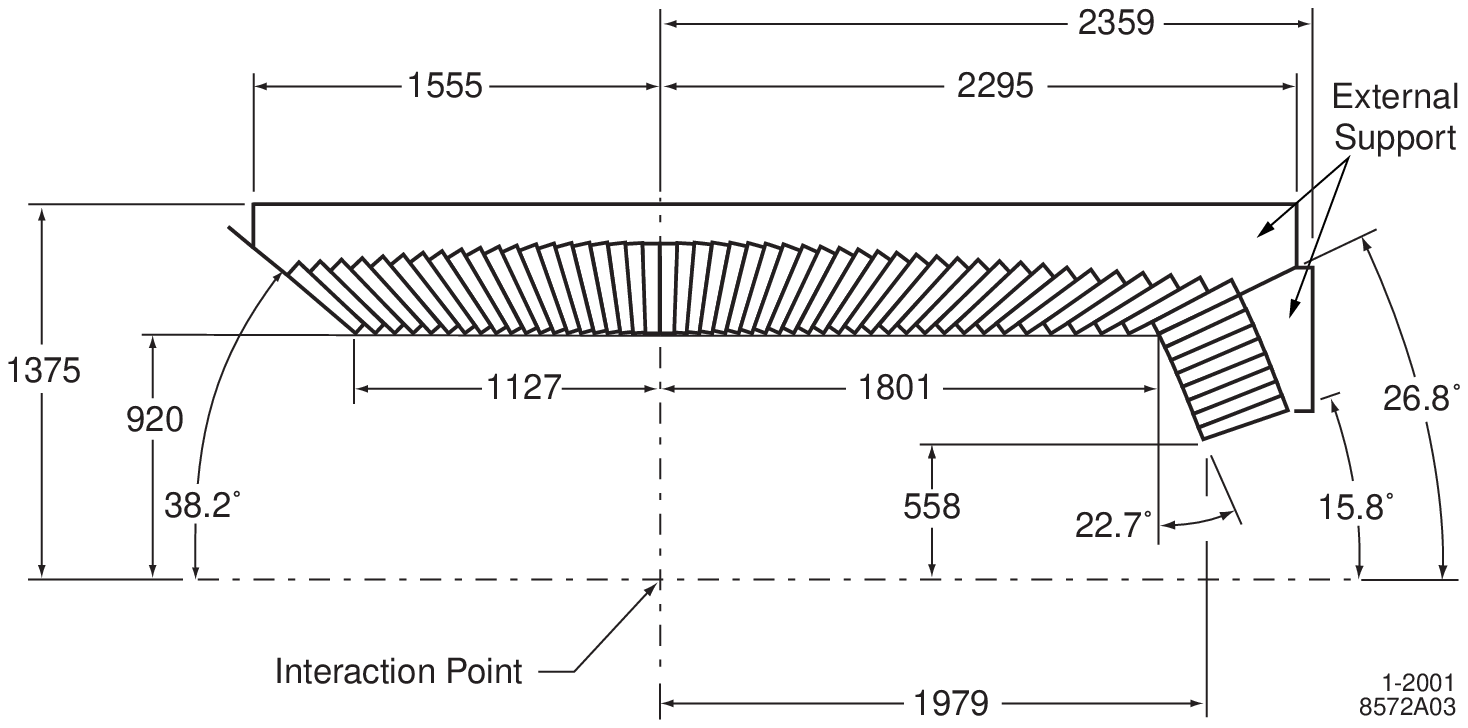}
\caption{Longitudinal cross section of the calorimeter.  Spatial 
dimensions are given in millimeters.}\label{fig:8572A03}
\end{figure}

\PARstart{S}{ince} 1999, the {\abf} at the Stanford Linear Accelerator
Center~(SLAC) has been collecting data from collisions of 9\GeV
electrons and 3.1\GeV positrons.  The~\B~mesons that are created at this
10.58\GeV center-of-mass energy are the major topic of the physics
studies at the \babar detector~\cite{bib:NIM} which records the decay
products of the \B~mesons.  The detection of photons and \piz mesons,
which mainly decay into photon pairs, with high efficiency and good
energy resolution is of utmost importance to the physics
program~\cite{bib:pi0pi0}~\cite{bib:newmeson}.  This~task is
handled by the electro-magnetic calorimeter~(EMC) of the \babar
detector, designed for the detection of photons in the energy range from
20\MeV to 8\GeV.  The detector consists of 5760~CsI(Tl) crystals
arranged cylindrically in 48 rows along the polar angle (the so-called
``barrel''), and 820~CsI(Tl) crystals arranged in 8~rows at the forward
end of the detector (``endcap''), as illustrated in
Figs.~\ref{fig:gv_emccut} and~\ref{fig:8572A03}.  Each crystal is of
trapezoidal shape with a front face area of approximately
47\mmeter$\times47$\mmeter and a length of 297.6\ensuremath{\,}mm to
325.5\ensuremath{\,}mm (16.0~to 17.5 radiation lengths).

\section{Requirements}

In order to reach good energy resolution in the calorimeter, accurate
calibration of the crystals over the full energy range is essential.
Because radiation from the \epem beams damages the crystals over time,
frequent calibrations are necessary to avoid a drop in energy
resolution.  For~lowest energies, this calibration is achieved through
the neutron-activated liquid-source system that is the subject of
this report.  A~calibration accuracy of 0.5\% was required
for the liquid source calibration system.  For comparison, we list here
the measured energy resolution of the calorimeter (first error
statistical, second error systematic)~\cite{bib:martin}:
\begin{equation}
 \frac{\sigma_E}{E} =
 \frac{(2.30\pm0.03\pm0.3)\%}{\sqrt[4]{E(\hbox{GeV})} } \oplus
 (1.35\pm0.08\pm0.2)\%.
\label{eq:reso}
\end{equation}

\section{Apparatus and Calibration Procedure}

\subsection{The Neutron Generator}

\begin{figure}
\centering \includegraphics[width=3.5in]{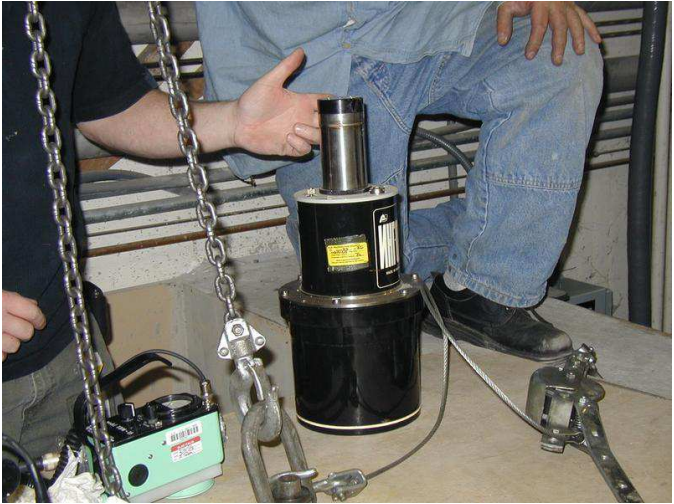}
\caption{The~neutron-generating unit of the neutron generator
system.}\label{fig:PB140023.med.eps}
\end{figure}

\begin{figure}
\centering
\includegraphics[width=3.5in]{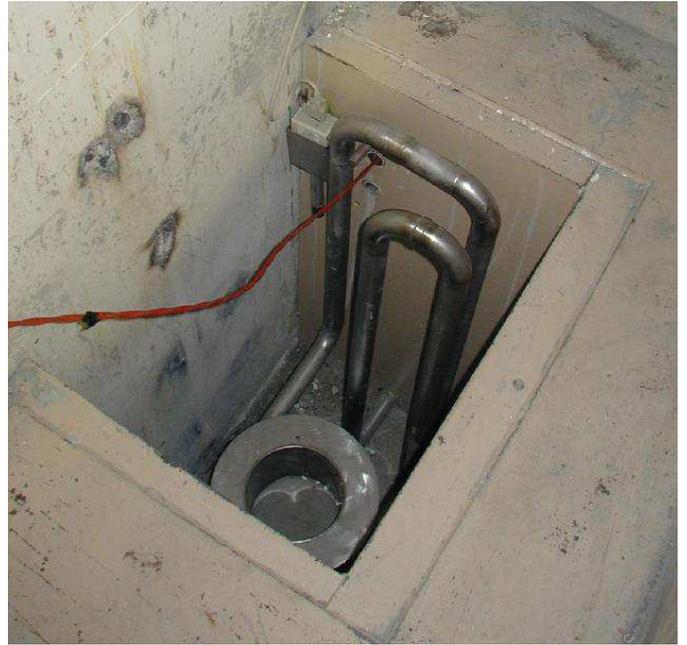}
\caption{A~look into the radiation shielding bunker.  The~round
structure inside is the bath designed to hold the neutron
generator.}\label{fig:PB140030_crop.eps}
\end{figure}

The calibration relies on the reaction $^{19}{\rm F} + \hbox{n} \to
\hbox{$^{16}$N}+\alpha$ and the subsequent decay of $^{16}{\rm N}$
($T_{1/2}=7\,\hbox{seconds}$) via $^{16}$O$^*$ into $^{16}{\rm O}$ and a
6.13\MeV photon.  The~neutrons originate from a~deuterium-tritium
neutron generator (\cite{bib:VNIIA}, Fig.~\ref{fig:PB140023.med.eps})
located in a concrete bunker with approximately 0.8\meter thick
walls adjacent to the \babar detector.  It~provides 14\MeV neutrons with
a continuous flux of up to $10^9$~n/s when its high voltage is turned on
remotely.  A~bath (Fig.~\ref{fig:PB140030_crop.eps}) surrounding the
generator holds Fluorinert{\texttrademark} FC-77~\cite{bib:fc77}, an
inert fluid rich in fluorine that the neutrons activate to~$^{16}{\rm
N}$.

\subsection{The Tube System}

\begin{figure}
\centering \includegraphics[width=3.5in]{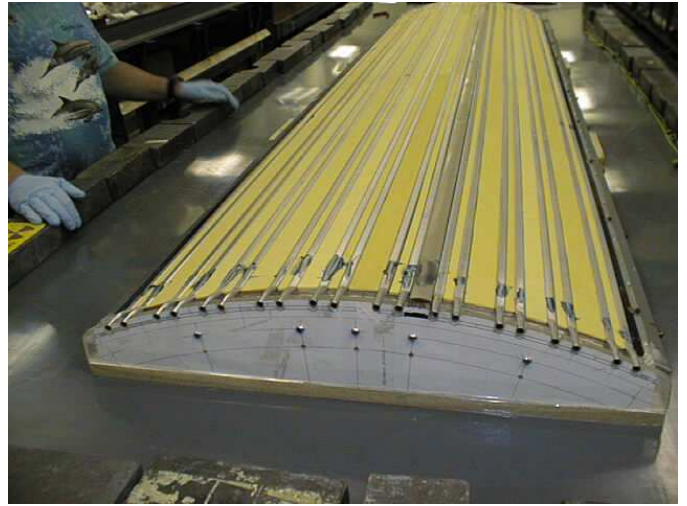}
\caption{Fabrication of the barrel panels with the top aluminum shield
still missing.  One can see the \y{f}\null\y{l}attened aluminum tubes running up and
down embedded in polyurethane.}\label{fig:PIC00009.eps}
\end{figure}

\begin{figure}
\centering
\includegraphics[width=3.5in]{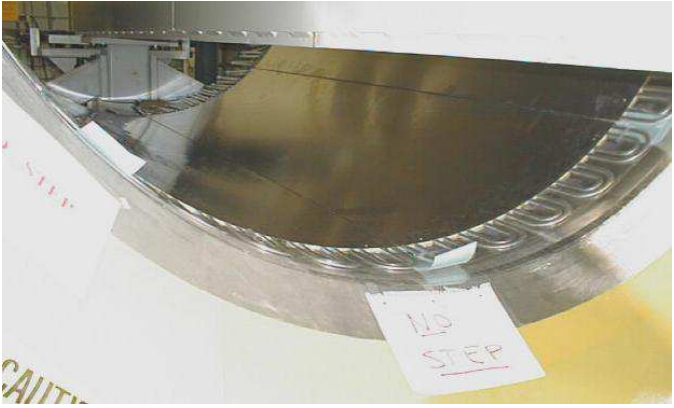}
\caption{Barrel panels in place at the inside of the
electromagnetic calorimeter.}\label{fig:panel_inst.eps}
\end{figure}

\begin{figure}
\centering
\includegraphics[width=3.5in]{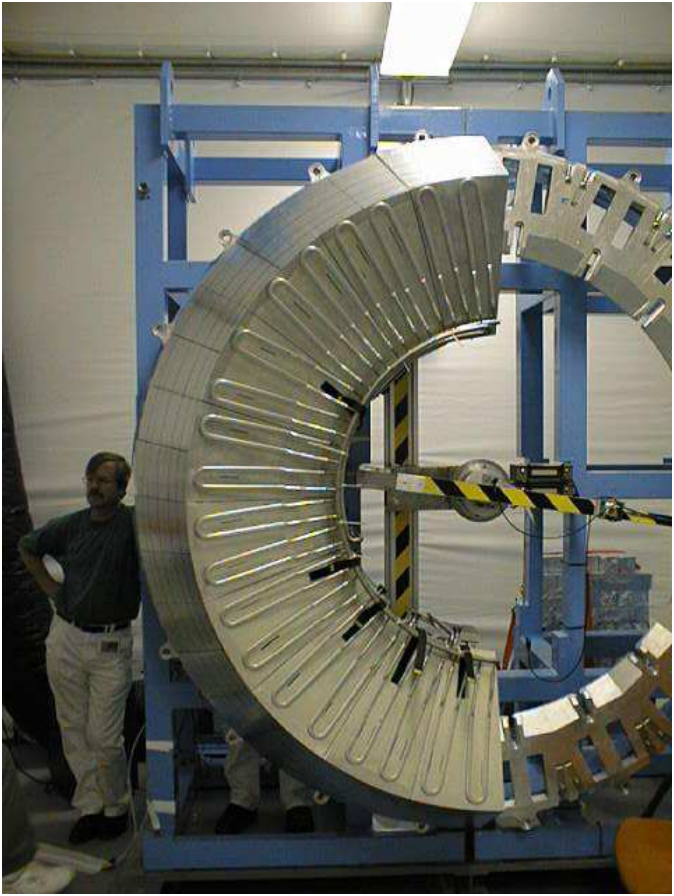}
\caption{One half of the endcap with installed network of \y{f}lattened
aluminum tubes in the front.}\label{fig:pic00019.eps}
\end{figure}

A~pump with a supply pressure of around $4\,$kPa transports the
activated liquid at a rate of 150 L/min, from the neutron generator and
bath, through pipes into the \babar detector, forces it through
thin-walled aluminum tubes in front of the crystals, and leads it back
to the bath in a continuous loop.  Since only a 15\mmeter gap was
available between the inside of the calorimeter and the next subsystem,
the panels sliding into this gap were allowed to have a thickness of
maximal 5\mmeter.  Fig.~\ref{fig:PIC00009.eps} shows one such panel in
production.  Round aluminum tubes of 10\mmeter diameter with
0.5\mmeter wall thickness were flattened to a 3\mmeter height, embedded
in polyurethane and covered with 0.4\mmeter thick aluminum sheets on
both sides.  Fig.~\ref{fig:panel_inst.eps} shows barrel panels at their
installation location at the inside of the calorimeter.  The design was
different for the endcap, as can be seen in Fig.~\ref{fig:pic00019.eps}.

\subsection{The Data Acquisition}

Photons emitted in the decay of $^{16}{\rm N}$ enter the crystals of the
calorimeter at a typical rate of 40\hz per crystal.  According to
Monte Carlo simulations, about 30\% of them deposit at least 3\MeV in an
individual crystal.  The energies are measured through the regular data
acquisition system, starting with the detection of the scintillation
light by a pair of photo diodes.  The~signal from the photo diodes
passes through an analog amplifier mounted at the back of the crystal,
then to a 3.7\mhz digitizer at the side of the detector.  Via optical
fiber the signal arrives in Versa Module Europa (VME) crates at
readout modules (ROMs) with embedded processors running VxWorks.  Since
the data acquisition system is designed for the triggered collection of
colliding beam events, not for random photons from the source
calibration system, a~pulse generator provides triggers of about
20$\,$kHz to begin the data acquisition sequence.

\begin{figure}
\centering
\includegraphics[width=2.8in]{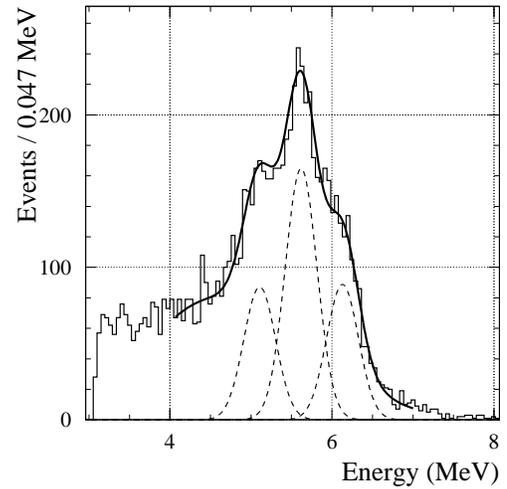}\vspace{-7mm}
\caption{Calibration spectrum of one crystal.  The \y{f}it identi\y{f}ies
  the contribution of the 6.13\MeV photons (right Gaussian) and the
  1$^{\rm st}$ and 2$^{\rm nd}$ escape peak (middle and left
  Gaussians).}\label{fig:bestfit}
\end{figure}

The ROMs select 128~samples of the wave form that arrives from each
crystal.  Since the decay time of CsI(Tl) is quite long (several hundred
nanoseconds) and since the samples are taken every 270\nsec, the signal
spreads over several samples of the wave form.  If~at least one of these
samples is above a pre-defined threshold of the crystal, the wave form
is sent to a~digital filter.  The filter is based on weights optimized
for the background conditions of the source calibration and is able to
reduce the electronics noise.  Then~the peak height is determined with a
parabolic fit to the peak of the filtered wave form.  If~the peak is
still above the threshold of the crystal, its value in ADC counts is
accumulated into an energy spectrum as shown in
Fig.~\ref{fig:bestfit}.  At~the end of the source calibration run, each
ROM writes the spectra to a file.

Since data are taken at a trigger rate of about 20\khz and since each
wave form has an effective time window of 26\usec, a live-time of about
50\% is reached.  Noisy channels can decrease the performance of the
data acquisition, and they are therefore masked out as well as
possible by raising their thresholds for the data acquisition.

\subsection{Fitting the Energy Spectra}

\begin{figure}
\centering
\includegraphics[width=2.7in]{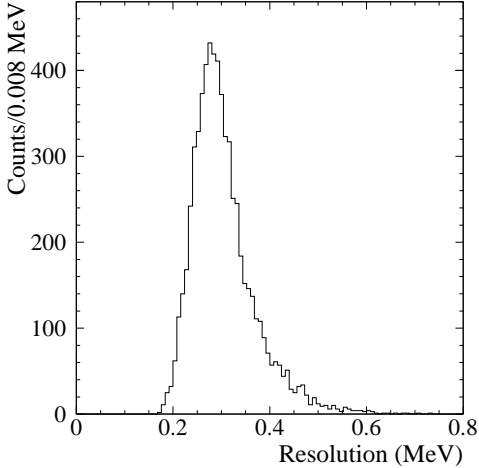}
\caption{Measured resolutions for a typical source calibration run.
The~resolution is de\y{f}ined as the width of each of the three
Gaussians illustrated in Fig.~\ref{fig:bestfit}.}\label{fig:reso}
\end{figure}

The energy spectra are fitted off-line to three Gaussians centered
around the 6.13\MeV peak and two escape peaks and to a parameterization
of the Compton background and electronics noise.  Other lines from the
decay of $^{16}$N are negligible at the present energy resolution.
Since the energy of the peaks is well-known, the fit returns the overall
gain (number of ADC counts per \MeVnospace) as well as the resolution,
defined here as the widths of each of the three Gaussian peaks.  Since
digital filtering reduces the electronics noise to about 230\KeV, an
average energy resolution of about 300\KeV is
reached~(Fig.~\ref{fig:reso}).  This resolution is better than
expected from (\ref{eq:reso}) because no beams are present during the
source calibration, the digital filter weights are optimized for this
no-beam situation, and because the photons originate right in front of
the crystals and reach the crystals after passing through only a
small amount of material.  The systematic error of the measurement is
estimated to be not more than~0.1\%.

\subsection{Applying the Constants in Data Analysis}

After each calibration run, the~calibration constants of each crystal
are recorded into a database.  When data collected during normal \epem
collisions are processed, the energy measured in each crystal is
adjusted via an interpolation between the low-energy source calibration
constants and high-energy calibration constants from Bhabha
events~\cite{bib:NIM}.

\section{Operational and Safety Issues}

The liquid source system is able to measure the light yield to high
precision in runs that last 15 to 30~minutes.  The~runs are performed
every ten days, on average, at times when no \ep or \en beams are
present in the collider rings.  The~aluminum panels with flattened tubes
and liquid add only about 2\% in radiation length to the material in
front of the calorimeter and therefore affect the physics data in only a
minimal way.  The~system requires nearly no maintenance, and safety
hazards are minimal.  The neutron generator is located safely inside a
locked radiation-shielding bunker and only generates neutrons when its
high voltage is turned on.  The hazard of Fluorinert{\texttrademark} is
negligible, even to the crystals, and the short life-time of the
activated fluid minimizes any radiation exposure in case of an
accidental spill.

\section{Results}

\begin{figure}
\centerline{\includegraphics[width=3.8in]{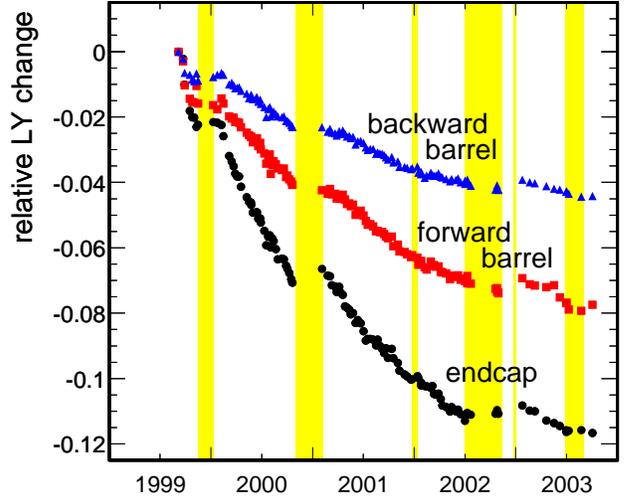}}
\vspace{-10mm} 
\caption{Relative change in light yield (measured as gain in ADC
  counts/\MeVnospace) since September 1999.  The average shift of the
  crystals is shown for three different subsets of the crystals.
  The~shaded areas indicate major periods without
  beams.}\label{fig:gainhistory}
\end{figure}

Calibration runs started in 1999 when the {\abf} began its operation.
Fig.~\ref{fig:gainhistory} illustrates how the average light yield of
the crystals dropped over time for the endcap, the forward barrel, and
the backward barrel.  The largest drop is seen in the forward endcap
($\sim12\%$), while the smallest drop is found in the backward barrel
($\sim4\%$).  The shaded areas in Fig.~\ref{fig:gainhistory} mark major
durations without beams.

\begin{figure}
\centering \includegraphics[width=3.5in]{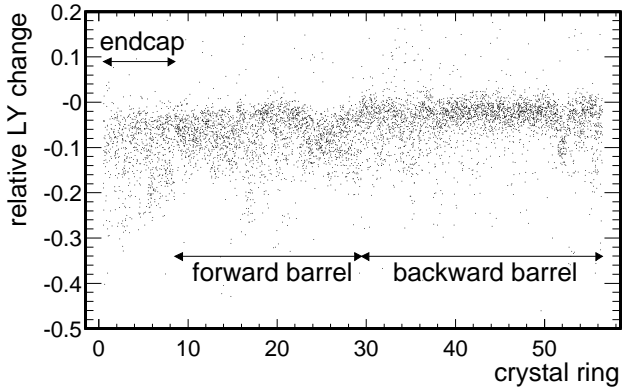}
\vspace{-6mm}
\caption{Distribution of the relative light yield change versus
crystal rings.}\label{fig:d}
\end{figure}

\begin{figure}
\centering \includegraphics[width=3.5in]{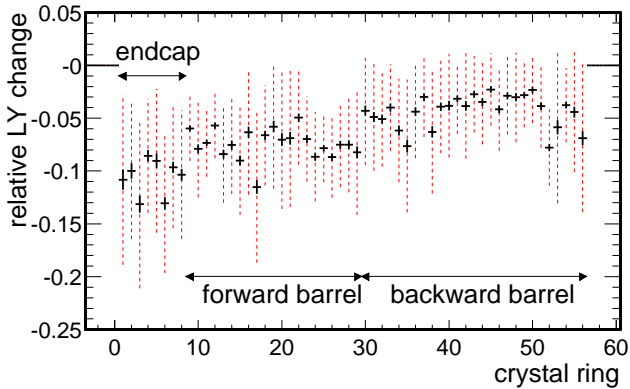}
\vspace{-6mm}
\caption{\Black{Relative change in light yield from September 1999 to
October 2003 versus crystal ring.  The solid error bars indicate the
error on the mean, the dotted line the spread.  Crystals with light
yield changes above 5\% and below $-$25\% were excluded to avoid bias
through outliers.}}\label{fig:f}
\end{figure}

Each point in Fig.~\ref{fig:gainhistory} is an average over
crystals from different manufacturers, located in different places within
the calorimeter, and receiving different radiation doses.  To illustrate
the spread of these measurements, Fig.~\ref{fig:d} presents the
light yield change from September 1999 to October 2003 versus the
56 crystal ring.  The crystal ring number corresponds to the order in
polar angle as shown in, e.g., Fig.~\ref{fig:8572A03}.  The endcap
ranges from ring 1 to ring 8, which is adjacent to the forward barrel
(rings 9 to 29), followed by the backward barrel (rings 30
to 56).  In~the average, a~larger change is seen toward the forward end
(low crystal ring numbers), and smaller changes at the backward end, but
within each crystal ring the crystals show quite some spread.  The~same
data are plotted in Fig.~\ref{fig:f} with the mean and its error (solid
line) and the spread (dotted line).  One clearly sees several
rings with larger light yield changes than their neighbors, like crystal
rings 23 to 27 and ring 52.  Since those crystals were obtained from a
different manufacturer than the surrounding crystals, the dependency of
the light yield change on the crystal manufacturer is studied more
below.

\begin{figure}\Black{
\centering 
\includegraphics[width=1.25in]{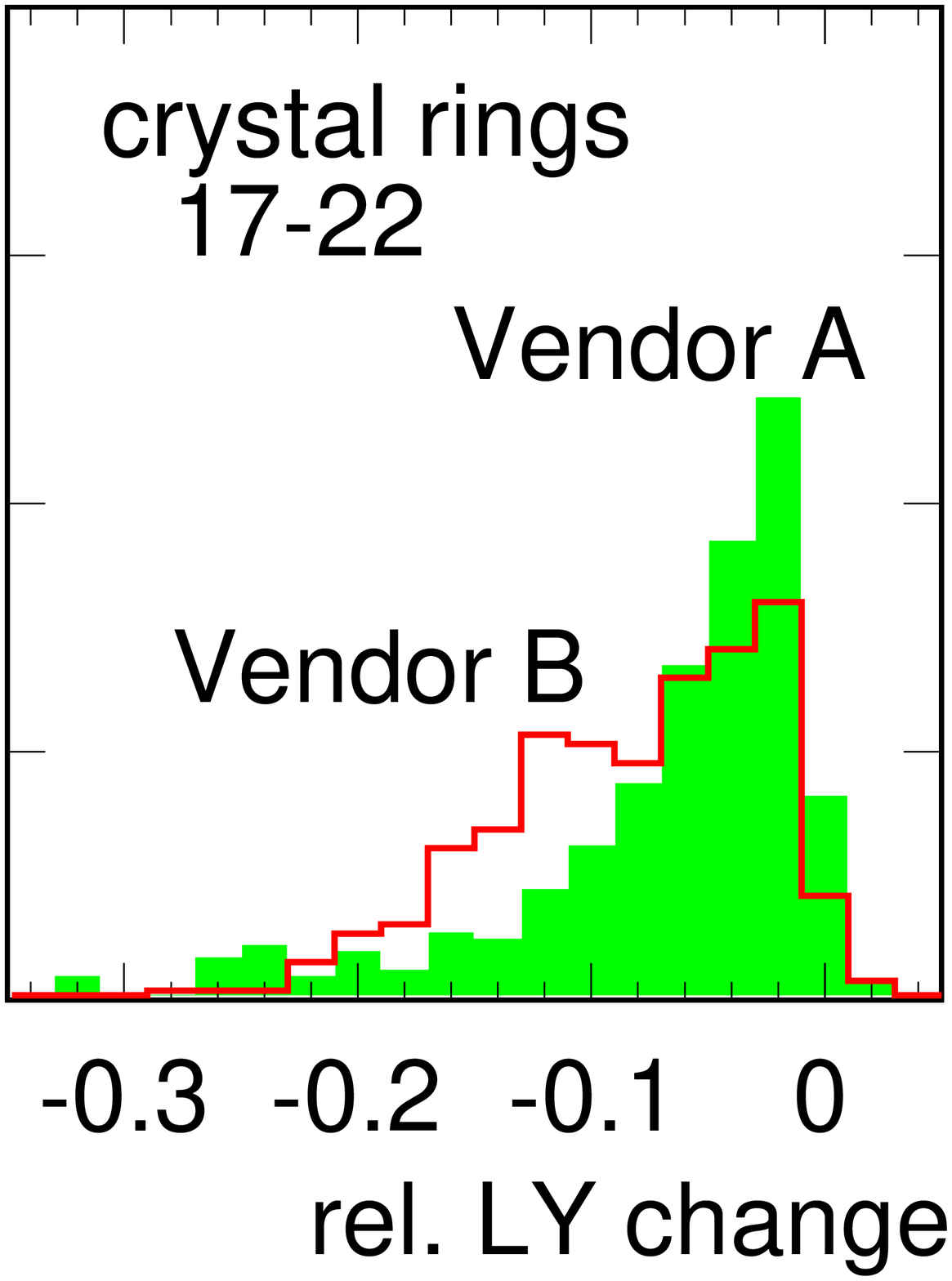}\hskip-3mm
\includegraphics[width=1.25in]{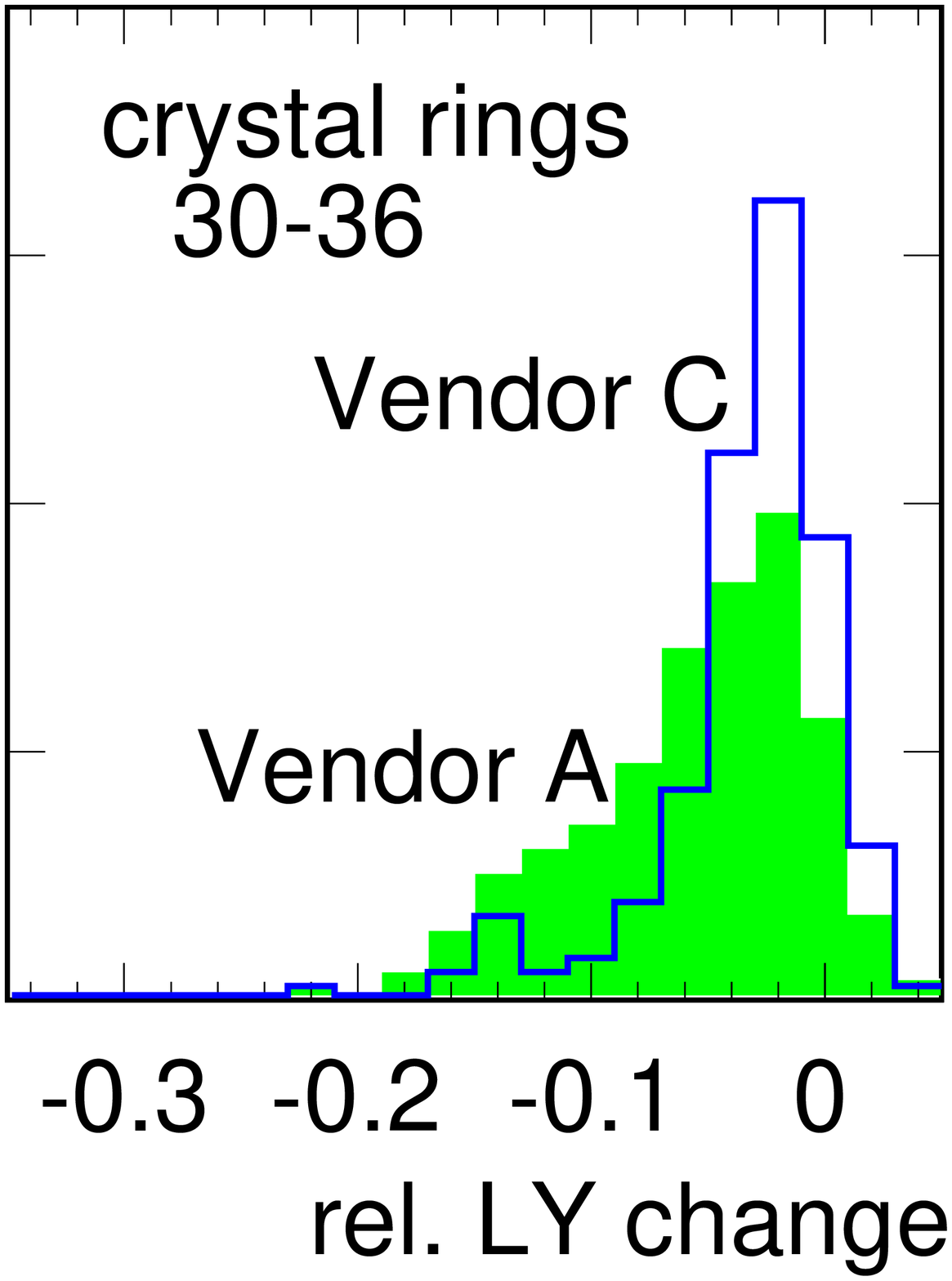}\hskip-3mm
\includegraphics[width=1.25in]{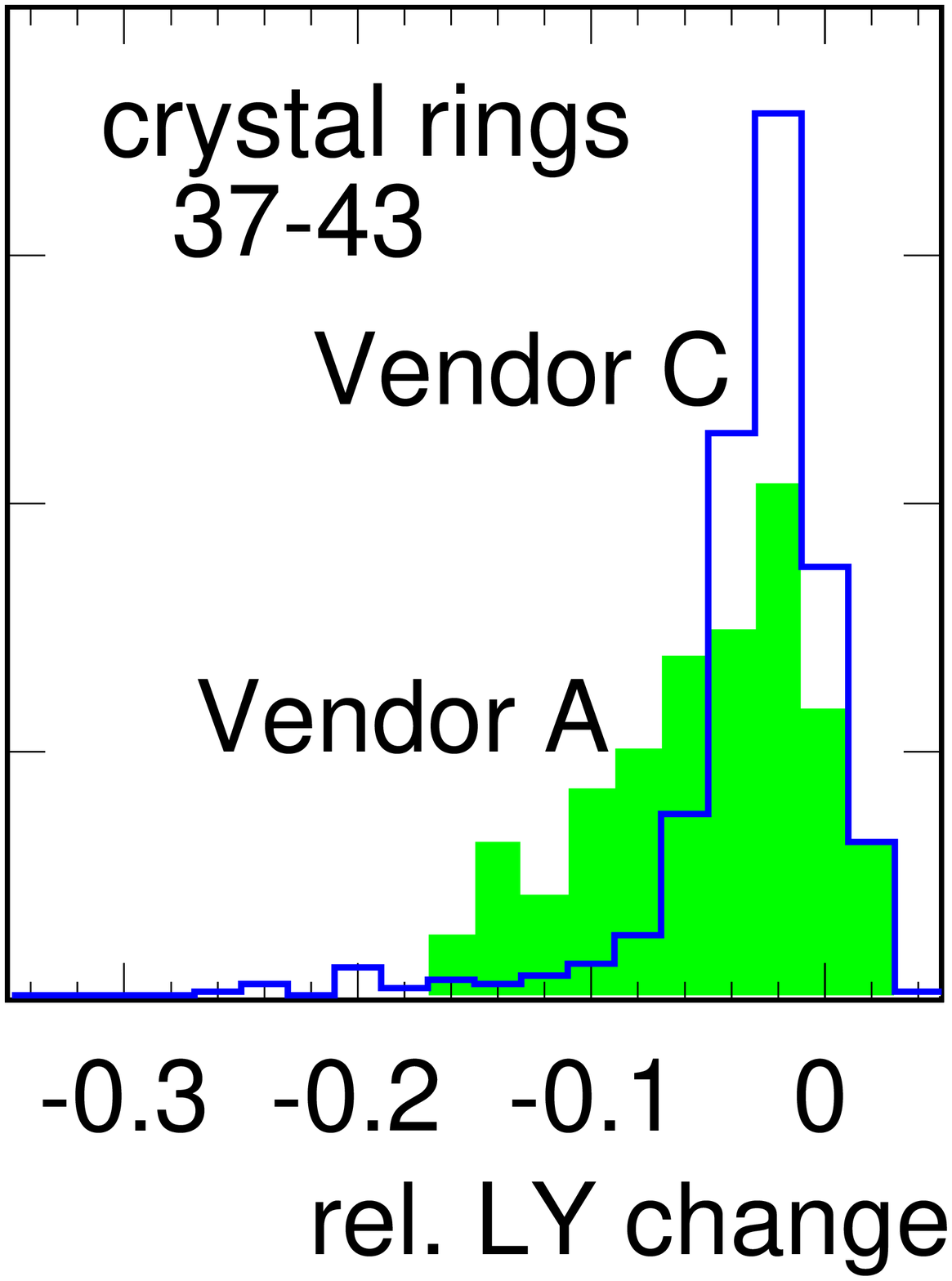}
\vspace{-8mm}
\caption{Relative change in light yield from September 1999 to
  October 2003 for crystals in three regions of crystal rings.  The
  solid histogram contains the numbers for Vendor~A, while the open
  histogram contains the numbers for Vendor~B (left) and Vendor~C
  (middle \& right).  The open and solid histograms are normalized
  to the same area.}\label{fig:someVendorChange}}
\end{figure}

Each plot in Fig.~\ref{fig:someVendorChange} shows the distribution of
the light yield change for a certain range of crystal rings for two
manufacturers.  Clearly the light yield change is less for the solid
histogram in the left plot (Vendor~A) than for the open histogram
(Vendor~B).  In~the next two plots, the crystals from Vendor~C (open
histograms) show even less change than the crystals from Vendor~A.
By~restricting each plot in Fig.~\ref{fig:someVendorChange} to few
crystal rings, the crystal that are compared are ensured to have
received similar radiation doses.

\begin{figure}
\Black{\centerline{\includegraphics[width=3.8in]{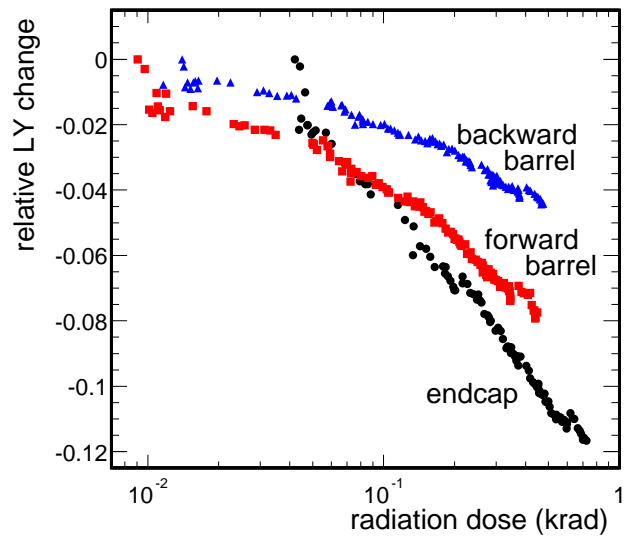}}
\vspace{-6mm} 
\caption{Relative change in light yield since September 1999 (as
shown in Fig.~\ref{fig:gainhistory}) plotted versus the radiation dose
measured by RadFETs.}}\label{fig:RadFET}
\end{figure}

The radiation received by the calorimeter is also monitored by 116
special Field Effect Transistors (RadFET) mounted at the front of the
crystals~\cite{bib:radfet}.  Although their measurements might not
be completely representative for the dose received by the crystals
because they are based on different physical principles, important
conclusions may still be drawn.  The measurements indicate that the
radiation dose inside the barrel is not varying widely, so that the
difference in the light yield changes seen from the source calibration
runs might be mainly due to crystals originating from different vendors.
The radiation dose to the endcap, however, is larger than to the barrel.
The endcap is thought to receive additional radiation at the rear
of the crystals from beam background originating from the beam line, and
not only radiation from the front.  Similar conclusions on the
radiation dose are also reached from studies of the change in the
leakage current of the crystals' photo
diodes~\cite{bib:tetianaCalor2002}.  In Fig.~\ref{fig:RadFET}, the
light yield changes are plotted versus the radiation dose as measured by
the RadFET at the time of the source calibration runs.  Since RadFET
measurements are available from May 1999 onward, when beams started to
pass through the \babar detector, the data points in
Fig.~\ref{fig:RadFET} are offset horizontally, especially for the
endcap.  Many factors may affect the quantities shown in this plot, for
example, the beam conditions that were changing over time, and care
must be taken when interpreting the plot.

At the time of construction, the~specifications for the crystals
required that their light yield response was uniform along the length of
the crystal.  But since during operation many more low-energy photons
hit the crystals in the front than in the rear (especially in the
barrel), it is possible that the front of the crystals is being damaged
more than the rear of the crystals.  So~far the drop in light yield
appears to be similar when comparing the results from the low-energy
liquid source calibration and the high-energy Bhabha
calibration~\cite{bib:tetianaCalor2002}.  Changes in light yield
uniformity are further studied in~\cite{bib:tetiana}.

\section{Conclusion}

Operating with minimal maintenance requirements and minimal safety
hazards, the liquid source calibration system at \babar is frequently
measuring the light yield of the crystals at low energy to the 0.5\%
accuracy that it was designed for.  By~monitoring the light yield
changes, the system corrects for the effects of radiation damage
caused to the crystals over time and allows the resolution of the
calorimeter to remain optimal.  The liquid source system therefore
serves the need of the electro-magnetic calorimeter very well.

\section*{Acknowledgment}
The author would like to thank the \babar EMC group for all their
contributions to the calorimeter and the liquid source system, making
possible and very worthwhile the calibrations described in this report.
He~would also like to congratulate and extend his gratitude to the whole
\babar Collaboration and the PEP-II accelerator group for their
tremendous accomplishments.

\end{document}